\documentclass[aps,pra,amsmath,amssymb,amsfonts,reprint,groupedaddress,showpacs,showkeys,floatfix]{revtex4-1}

\usepackage[utf8]{inputenc}
\usepackage{siunitx}
\usepackage{graphicx}
\usepackage{revsymb4-1}
\usepackage{bm}
\usepackage{times}
\usepackage{soul}
\usepackage{ulem}

\usepackage[usenames,dvipsnames]{xcolor}

\newcommand{\eg}{e.g.\ }
\newcommand{\figref}[2][]{Fig.~\ref{fig:#2}#1}
\newcommand{\eqnref}[1]{Eq.~(\ref{eq:#1})}
\newcommand{\secref}[1]{{Sec.~(\ref{sec:#1})}}

\newcommand{\tabref}[1]{table (\ref{tab:#1})}
\newcommand{\ybgion}{\textsuperscript{172}Yb\textsuperscript{+}}
\newcommand{\matr}[1]{{\bm{#1}}}
\newcommand{\mean}[1]{\left\langle\,#1\,\right\rangle}
\renewcommand{\vec}[1]{{\bm{#1}}}
\newcommand{\unitvec}[1]{{\bm{\hat{#1}}}}

\begin{document}

\normalem


\title{Ion trajectory analysis for micromotion minimization and the measurement of small forces}


\author{Timm~F.~Gloger}
\author{Peter~Kaufmann}
\author{Delia~Kaufmann}
\author{M.~Tanveer~Baig}
\author{Thomas~Collath}
\author{Michael~Johanning}
\author{Christof~Wunderlich}

\email{wunderlich@physik.uni-siegen.de}
\homepage{http://quantenoptik.uni-siegen.de}
\affiliation{Department Physik, Naturwissenschaftlich-Technische Fakultät, Universität Siegen, 57068 Siegen, Germany}


\date{October 27, 2015}

\begin{abstract}
For experiments with ions confined in a Paul trap, minimization of micromotion is often essential.
In order to diagnose and compensate micromotion we have implemented a method that allows for finding the position of the radio-frequency (rf) null reliably and efficiently, in principle, without any variation of direct current (dc) voltages. We apply a trap modulation technique and focus-scanning imaging to extract three-dimensional ion positions for various rf drive powers and analyze the power dependence of the equilibrium position of the trapped ion. In contrast to commonly used methods, the search algorithm directly makes use of a physical effect as opposed to efficient numerical minimization in a high-dimensional parameter space. Using this method we achieve a compensation of the residual electric field that causes excess micromotion in the radial plane of a linear Paul trap  down to \SI{0.09}{\volt\per\meter}. Additionally, the precise position determination of a single harmonically trapped ion employed here can also be utilized for the detection of small forces. This is demonstrated by determining light pressure forces with a precision of \SI{135}{\yocto\newton}. As the method is based on imaging only, it can be applied to several ions simultaneously and is independent of laser direction and thus well-suited to be used with, for example, surface-electrode traps.
\end{abstract}


\maketitle

\section{Introduction}\label{sec:Introduction}

Trapped laser-cooled ions are a prolific starting point for many experiments related to quantum information science \cite{Wineland1998B,Blatt2008} and precision spectroscopy, yielding some of the most accurate clock standards to date \cite{Chou2010, Huntemann2012, Barwood2012}. Ions can be trapped for long times and laser-cooled, even to the motional ground state \cite{Diedrich1989, Eschner2003, Segal2014}. They are  one of the most promising candidates for quantum computation \cite{Blatt2008} and for quantum simulations where the quantum mechanical properties of a system difficult to investigate directly is simulated using the well-understood and -controlled quantum system of trapped ions \cite{Friedenauer2008,Johanning2009,Kim2010,Lanyon2011,Blatt2012,Schneider2012}.
The size of this quantum system can be scaled up by entangling few-ion systems in complex segmented trapping architectures  \cite{Kielpinski2002, Hughes2011, Monroe2014}.
	
The conceptual starting point for all this research is one or several ions at rest, more accurately, close to the motional ground state of an effective, approximately harmonic trapping potential \cite{Dehmelt1967, Paul1990}.
The Laplace equation forbids the existence of electrostatic potential minima in vacuum (\emph{Earnshaw theorem}). This limitation is bypassed in radio-frequency (rf) quadrupole (Paul) traps, where the ions are confined by an inhomogeneous oscillating field \cite{Dehmelt1967, Paul1990}, and the oscillation energy of the forced oscillation becomes position dependent. Upon the presence of a cooling mechanism such as laser cooling, an ion comes to rest at the minimum of the oscillating field amplitude.
Additional direct current (dc) fields from biased electrodes or surface charges, \eg, created by the loading process, can add forces which push the ion away from the rf null; in addition, a phase mismatch between rf electrodes might even prevent the existence of a time-independent rf null. In both cases, the ion will carry out a forced oscillation at the drive frequency, the so called micromotion, with an amplitude depending on dc and phase mismatch.
Micromotion plays an important role in excess heating of trapped ions, \eg, through unwanted Doppler shifts \cite{Bluemel1989,DeVoe1989,Peik1999,Cirac1994} and in conjunction with inevitable anharmonicities or stochastic changes of the trapping potential \cite{Gudjons1997,Brouard2001,Ryjkov2005}.

Renewed interest in micromotion minimization comes from recent research with combined traps of neutral and charged atoms \cite{Grier2009,Zipkes2010, Schmid2010} where excess micromotion is a detrimental source of neutral atom loss and has prevented collisional cooling of ions by a cold neutral atom cloud to the fundamental limits \cite{Cetina2012}. Here, as an initial stage of combined traps, micromotion is carefully compensated and the neutral atom loss can be used as a figure of merit for optimization \cite{Haerter2013}. Micromotion can be overcome altogether by using a dipole trap for the ion. As the light interacts with the dipole moment of the ion instead of the charge, dipole traps are much softer and shallower and themselves require carefully balanced dc offset fields as a prerequisite \cite{Schneider2010}.
Another motivation for low micromotion comes from precision spectroscopy and frequency standards: micromotion sidebands spoil the accuracy of the determination of atomic resonance frequencies due to Doppler shifts, and thus micromotion minimization is an essential prerequisite for accurate ion trap frequency standards. One disadvantage, though not a matter of principle, is that ion clocks suffer in short term stability, as just one atom is probed. Massively parallel interrogation of many ions in common or separate traps, all micromotion compensated, is an important step to improve the short term stability of ion based frequency standards \cite{Pyka2014}.

Micromotion minimization can thus be regarded as a common crucial initial step and prerequisite for good trap performance.
Micromotion causes diverse signatures, for which several minimization methods have been developed:

\begin{enumerate}
	\item The ion's absorption and emission spectra are altered due to the periodic Doppler shifts caused by micromotion directly or by mechanisms conditioned on the presence of excess micromotion.
	
	Micromotion minimization is performed by exploiting the Doppler shift induced temporal correlation between the scattering rate and the rf trapping voltage \cite{Schubert1995,Berkeland1998,Pyka2014} and by measurement of micromotion sidebands in the absorption spectrum \cite{Berkeland1998, Schulz2008, Chwalla2009, Akerman2012, Warring2013} or the emission spectrum \cite{Chuah2013} spaced by the trap drive frequency.

For ion crystals the normal mode spectrum can be altered by excess micromotion and compensation of stray electric fields can be performed by minimization of these frequency shifts \cite{Barrett2003}.	
	\item\label{methodrevisisted} The ion's equilibrium position depends on the strength of the effective trapping potential if the dc fields are not nulled at the rf node.
	
	Excess micromotion is detected by monitoring the ion's average position while changing the strength of the effective trapping potential using either dc fields \cite{Schneider2005} or changing the rf voltage amplitude \cite{Berkeland1998}.
	\item Parametric resonances of the ion's motion can be excited by a modulation of the rf trapping voltage, if the dc fields are not nulled at the rf node.
	
	The excitation of such a resonance causes a change in the ion's scattering rate, which is then minimized \cite{Ibaraki2011,Narayanan2011,Tanaka2012}.
	\item Micromotion causes an collisional transfer of kinetic ernergy from the ion's micromotion to neutral atoms when trapped simultaneously.
	
	Micromotion minimization is performed by measurements of the loss rate of the atom trap \cite{Haerter2013}.	
\end{enumerate}

In this paper, we revisit the trap modulation technique, (\ref{methodrevisisted}). For different rf levels, we determine the three-dimensional (3D) ion position from a tomographic imaging procedure. This position can be determined with an uncertainty far below the wavelength of light  scattered off the ion for observation. 
From the ion positions measured at different rf powers, we can extrapolate the ion trajectory to infinite rf power, which ends at the rf null. 
With the knowledge gathered from a single trajectory, the ion can be moved to the rf null by changing the dc electric fields, which requires accurate \emph{a priori} characterization of the trapping fields. As an alternative, trajectories recorded at different dc settings readily yield this characterization, together with the optimized compensation voltages.

The time required for the minimization process can be crucial if the dc fields are fluctuating. This can happen as a consequence of stray charges and, usually to a lesser extent, by drifting fields from unstable voltage supplies \cite{Narayanan2011,Haerter2014}. For surface traps trapping times can be in the second or minute range owing to smaller trap depths and might require frequent reloading and patch potentials may change on a short timescale. In such cases it is desirable to carry out the minimization fast and repeatedly to track or compensate time-dependent effects.

The method discussed here relies on position determination, which can be carried out with a high precision and does not require any narrow transition and ultra stable lasers for sideband spectroscopy. A single position determination in two dimensions requires only milliseconds of measurement time, accumulating to a few hundred milliseconds of data acquisition for the complete minimization process. If position determination and the minimization are carried out in three dimensions, we need to use a tomographic method and the approach is slowed down by about an order of magnitude. On the other hand, even then, no constraints on the propagation direction of the laser limit the optimization procedure. As the position determination is based on imaging, the optimization might even be carried out for several ions in parallel, which is of particular interest for frequency standards. 

The method discussed here is fast, has a sensitivity comparable to that of other approaches and is yet general, simple, and economic. 

The paper is structured as follows: in Sec.~\ref{sec:Theory}, we model the displacement of the ion's equilibrium position. The experimental setup is detailed in Sec.~\ref{sec:experimental_setup}. The measurement procedure and results are presented in Sec.~\ref{sec:MethodAndResults}. As the position determination can be precise to the nanometer level, we use this for the measurement of small forces in the example of the light pressure in Sec.~\ref{sec:LightPressureMeasurement}. 

\section{Theory}
\label{sec:Theory}

This section starts with an overview of the mathematical treatment of a single ion confined in a Paul trap in the presence of an electric stray field, both by solving the respective Mathieu equations \cite{Berkeland1998} and by using the pseudopotential approach \cite{Dehmelt1967}. Readers familiar with this treatment might be referred to \eqnref{averageDisplacement} and \eqnref{equilibriumEffectiveSimple} as the main results and proceed from \eqnref{effectiveSimpleTranslated} for a discussion of the signature ion trajectories evolving from this models.

An ion (mass $m$, charge $Q$) trapped in a Paul trap (operated at an rf voltage $V \cos{\Omega t}$ and a dc voltage $U$) in the presence of an electric stray field obeys the inhomogeneous Mathieu equations \cite{Berkeland1998},
\begin{equation}
	\label{eq:mathieuEquations}
	\ddot{r}_k + [a_k + 2 q_k \cos(\Omega t)]\frac{\Omega^2}{4}r_k = \frac{Q \varepsilon_{{\textrm{stray}},k}}{m},
\end{equation}
where $\vec{r} = \sum r_k \unitvec{k}$ ($k=x,y,z$) is the ion position, $\unitvec{k} = \unitvec{x},\unitvec{y},\unitvec{z}$ are the Cartesian unit vectors, $a_k$ and $q_k$ are the commonly called trapping parameters, and the electric stray field $\vec{\varepsilon}_{{\textrm{stray}}} = \sum \varepsilon_{\textrm{stray},k} \unitvec{k}$ is approximated to be static and uniform over the small volume occupied by the ion trajectory.

This stray field might be a real physical field (\eg, originating from charged insulators near the trap), but also effects lifting ideal symmetry and displacing the dc origin with respect to the rf (\eg, misaligned trap electrodes or machining imperfections) as well as other homogeneous static forces (\eg, the light pressure force exerted by a cooling laser) can effectively be treated as contributions to the stray field.

Solving \eqnref{mathieuEquations} using the adiabatic approximation
($|a_k|, q_k^2 \ll 1$) yields the approximate motion of the ion \cite{Berkeland1998}
\begin{eqnarray}
	r_k(t) &\approx& [r_{\varepsilon,k} + r_{0,k} \cos(\omega_k t + \varphi_k)]\left(1 + \frac{q_k}{2}\cos(\Omega t)\right) \\
	\label{eq:mathieuTrajectory}
	&=& [r_{0,k} \cos(\omega_k t + \varphi_k)]\left(1 + \frac{q_k}{2}\cos(\Omega t)\right) \nonumber\\
	&& + r_{\varepsilon,k} + \frac{q_k r_{\varepsilon,k}}{2}\cos(\Omega t),
\end{eqnarray}
where $\vec{r}_0$ and $\varphi_k$ depend on the initial conditions and $\omega_k = \frac{\Omega}{2} \sqrt{a_k + q_k^2/2}$.
\eqnref{mathieuTrajectory} describes a motion characterized by two frequency components:
the secular motion component, oscillating at frequency $\omega_k$, and the micromotion component, oscillating at the much higher frequency $\Omega$ and much lower amplitude $q_k r_{0,k}/2$.
The secular motion can be reduced by cooling mechanisms as, \eg, laser cooling.
This reduces the micromotion amplitude as well, as it is proportional to the secular motion.
The last two terms in \eqnref{mathieuTrajectory} describe the effects of the additional electric field.
The field $\vec{\varepsilon}_{\textrm{stray}}$ shifts the average position of the ion out of the rf potential node to the position
\begin{equation}
	\label{eq:averageDisplacement}
	\vec{r}_{\varepsilon} \approx \frac{Q}{m} \sum\limits_{k} \frac{\varepsilon_{\textrm{stray},k}}{\omega_k^2}\unitvec{k},
\end{equation}
at which the rf electric field causes oscillations with amplitudes $ q_{k} r_{\varepsilon,k}/2$ -- the excess micromotion. This motion is a driven motion and thus cannot be significantly reduced by cooling techniques.

In the pseudo- or ponderomotive potential approach \cite{Dehmelt1967}, the motion of the ion due to the rf electric field is averaged over one period of micromotion and the so-called rf pseudopotential represents the kinetic energy due to micromotion,
\begin{equation}
	\label{eq:pseudopotential}
	\phi_{\textrm{rf}}^{\textrm{eff}}(\vec{r}) = \frac{m}{2} \sum\limits_{k}{\omega_{\textrm{rf},k}^2 r_k^2}, \qquad \omega_{\textrm{rf},k}^2 = \frac{\Omega^2}{4} \frac{q_k^2}{2}.
\end{equation}
The $\omega_{\textrm{rf},k}$ are called the rf trap frequencies and represent the rf contribution to the secular frequencies $\omega_k$.

The ion can then be treated as being confined in an effective potential that is the sum of the rf pseudopotential $\phi_{\textrm{rf}}^{\textrm{eff}}$ and all dc potential contributions [including dc electrodes ($\phi_{\textrm{dc},i}$) as well as stray potentials ($\phi_{\textrm{stray}}$)],
\begin{equation}
	\label{eq:effectivePotential}
	\phi_{\textrm{eff}}(\vec{r}) = \phi_{\textrm{rf}}^{\textrm{eff}}(\vec{r}) + Q\sum\limits_{i}{\phi_{\textrm{dc},i}(\vec{r})} + Q \phi_{\textrm{stray}}(\vec{r}),
\end{equation}
where $-\nabla \phi_{\textrm{stray}}(\vec{r}) = \vec{\varepsilon}_{\textrm{stray}}$.
Here the $\phi_{\textrm{dc},i}$ are taken to be ideal potentials in the aforementioned sense: contributions causing a shift of the dc origin relative to the rf node (that is, contributions linear in the ion's position) are absorbed into $\phi_\textrm{stray}$.

Near the center of the confining potential (\eg, along the axis of a linear trap) the sum of dc potentials can be approximated as a quadrupole potential:
\begin{eqnarray}
	\label{eq:dcQuadrupole}
	Q \sum\limits_{i}{\phi_{\textrm{dc},i}(\vec{r})} = \frac{m}{2} \sum\limits_{k}{\omega_{\textrm{dc},k}^2 r_k^2}.
\end{eqnarray} 
Similarly to the $\omega_{\textrm{rf},k}$, the $\omega_{\textrm{dc},k}$ are the dc trap frequencies at a specific setting of all dc trapping voltages ($\omega_{\textrm{dc},k}^2 = \Omega^2 a_k / 4$ in the case of an ideal linear trap).
For the sake of simplicity the principal axes of the rf and dc potentials have been chosen here to be parallel and equal to the Cartesian axes.

\begin{figure}
	\centering
		\includegraphics[width=\columnwidth]{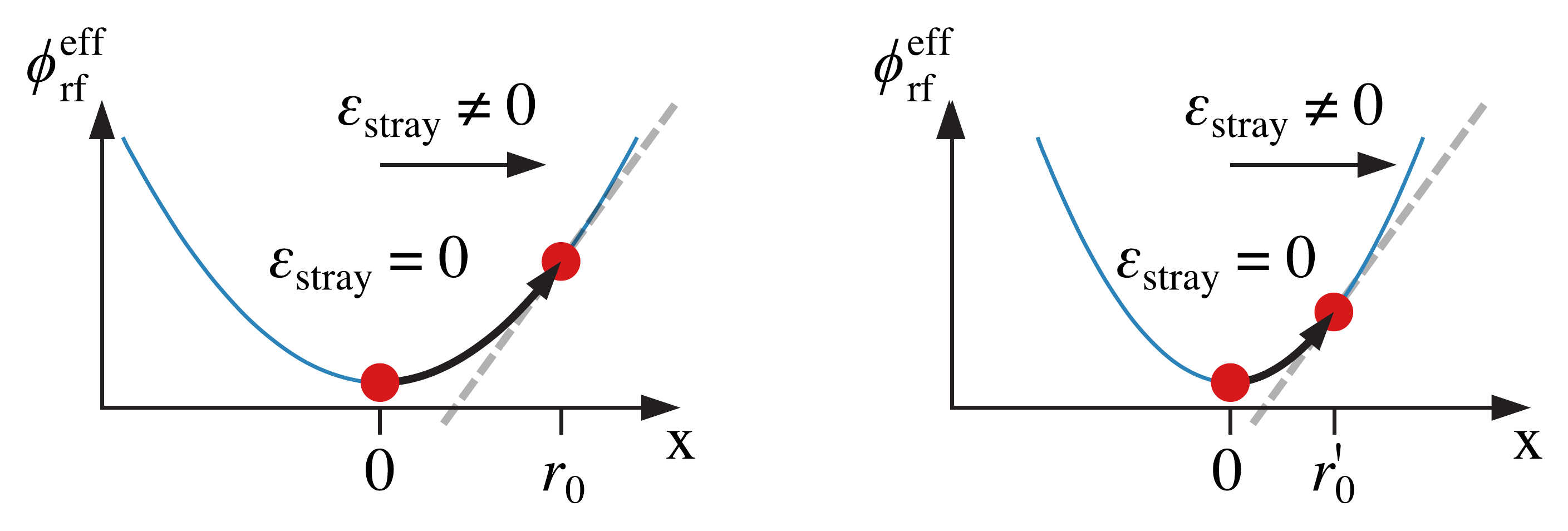}
	\caption{(Color online) Left: In the presence of a stray field $\vec{\varepsilon}_\textrm{stray}$, for example, an electric stray field from misaligned electrodes or the effective field of a light pressure force, the ion is pushed out of the minimum of the effective rf potential (thin blue line) to $r_\textrm{0}$ where it experiences a nonvanishing rf field and thus is subject to micromotion. Right: When the applied rf power is changed, the curvature of the effective rf potential and the equilibrium position $r^{\prime}_\textrm{0}$ of the ion change. As the stray fields in the left and right plots are of identical strength, the equilibrium condition as given in \eqnref{effectiveSimple} is fulfilled for identical slopes (dashed gray lines) of the effective rf potential at the ion's position.}
	\label{fig:concept}
\end{figure}

The equilibrium position of the ion is then given by zero net force
\begin{eqnarray}
\label{eq:effectiveSimple}
\vec{0} &\equiv& -\nabla \phi_{\textrm{eff}}(\vec{r}) \nonumber \\
&=& - \nabla \left[\frac{m}{2} \sum\limits_{k}{\left(\omega_{\textrm{rf},k}^2 + \omega_{\textrm{dc},k}^2 \right) r_k^2} + Q\phi_{\textrm{stray}}(\vec{r}) \right]\\
\label{eq:equilibriumEffectiveSimple}
	\Rightarrow \vec{r}_{\textrm{0}} &=& \frac{Q}{m}\sum\limits_{k} \frac{\varepsilon_{\textrm{stray},k}}{ \omega_{\textrm{rf},k}^2 + \omega_{\textrm{dc},k}^2} \unitvec{k} \nonumber\\
	 &=& \frac{Q}{m}\sum\limits_{k} \frac{\varepsilon_{\textrm{stray},k}}{ \frac{P_{\textrm{rf}}}{P_{\textrm{0}}}\omega_{\textrm{rf}_0,k}^2 + \omega_{\textrm{dc},k}^2} \unitvec{k},
\end{eqnarray}
which is equivalent to the average displacement $\vec{r}_{\varepsilon}$ from \eqnref{averageDisplacement}, as $\omega_k^2 = \omega_{\textrm{rf},k}^2 + \omega_{\textrm{dc},k}^2$.
Here $P_{\textrm{rf}}/P_{\textrm{0}}$ has been introduced as a convenient scaling factor. $P_{\textrm{rf}}$ is the applied rf power, and $\omega_{\textrm{rf}_0,k}$ the rf trap frequencies at an applied rf power of $P_{\textrm{0}}$ ($P \propto V^2 \propto q_k^2$).
The shift of the equilibrium position out of the rf potential minimum depends on the stray field and the applied rf power (see \figref{concept}).

As an intuitive point of view, it is also useful to combine the effects of all stray fields (real as well as effective fields) as a translation $\vec{\delta}(\vec{\varepsilon}_{\textrm{stray}})$ of the dc potential's origin explicitly and rewrite the effective potential as 
\begin{eqnarray}
	\phi_{\textrm{eff}}(\vec{r}) &=& \frac{m}{2} \sum\limits_{k} \frac{P_{\textrm{rf}}}{P_{\textrm{0}}}\omega_{\textrm{rf}_0,k}^2 r_k^2 \\
	&& + \frac{m}{2} \sum\limits_{k} \omega_{\textrm{dc},k}^2 \left[r_k - \delta_k(\vec{\varepsilon}_{\textrm{stray}})\right]^2. \label{eq:effectiveSimpleTranslated}
\end{eqnarray}

In a 1D trap, the electric stray field would shift the dc potential by
\begin{equation}
	\delta = -\frac{Q \varepsilon_{\textrm{stray}}}{m \omega_{\textrm{dc}}^2},
\end{equation}
and the ion's position in the trap by
\begin{equation}
	r_{\textrm{0}} = \frac{Q}{m} \frac{\varepsilon_{\textrm{stray}}}{ \frac{P_{\textrm{rf}}}{P_{\textrm{0}}}\omega_{\textrm{rf}_0}^2 + \omega_{\textrm{dc}}^2}.
\end{equation}

As the applied rf power is increased the ion's position will then approach the minimum of the rf pseudopotential ($r_{\textrm{0}} = 0$). By monitoring $r_{\textrm{0}}$ as a function of the rf power an extrapolation of the rf node in the limit of $P_{\textrm{rf}} \rightarrow \infty$ is possible. 

\begin{figure*}
\centering
\includegraphics[width=\textwidth]{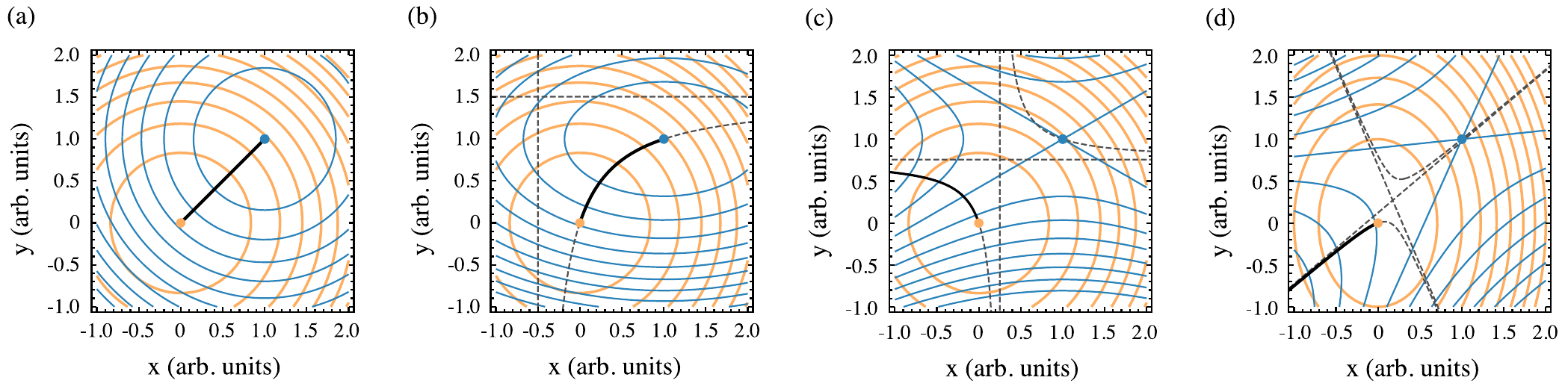}
\caption{(Color online) Two-dimensional potential configurations of the dc (blue isolines) and rf (pseudo-)potentials (thick yellow isolines). The potentials' origins are indicated by respectively colored filled circles. 
The thick black line represents the ion's trace as the strength of the rf potential is tuned from $\infty$ to 0. The dashed gray line shows the full hyperbola and its asymptotes describing the ion's movement.
(a) A configuration of degenerate trap frequencies ($\omega_{\textrm{rf},x}^2 = \omega_{\textrm{rf},y}^2 > 0$, $\omega_{\textrm{dc},x}^2 = \omega_{\textrm{dc},y}^2 > 0$).
Due to radial symmetry, the ion's equilibrium positions will move in a straight line.
(b) For broken dc radial symmetry ($\omega_{\textrm{dc},x}^2 < \omega_{\textrm{dc},y}^2$) the ion's trace changes to a hyperbola.
(c) A potential configuration with an unstable dc trapping in the $\unitvec{x}$ direction ($\omega_{\textrm{dc},x}^2 < 0$) as realized in the transverse plane of a linear Paul trap.
(d) Additionally broken rf radial symmetry ($\omega_{\textrm{rf},x}^2 > \omega_{\textrm{rf},y}^2$) and a rotation of the dc potential's principal axes causes a shearing and rotation of the hyperbolic trace, featuring nonperpendicular asymptotes.}
\label{fig:2d_potential_configs}
\end{figure*}

Instructive cases of 2D potential configurations displaying characteristic features of the ion's movement are shown in \figref{2d_potential_configs}. The electric stray potential has been included as a shift of the dc potential's origin [see \eqnref{effectiveSimpleTranslated}]. 
Rf potential isolines are drawn in yellow; dc potential isolines, in blue. Filled colored circles indicate the respective potential's origin. The thick black line marks the equilibrium positions of the ion while tuning the applied rf power from infinity to zero.

\figref[a]{2d_potential_configs} displays a configuration of confining rf and dc potentials with degenerate trap frequencies ($\omega_{\textrm{rf},x}^2 = \omega_{\textrm{rf},y}^2 > 0$, $\omega_{\textrm{dc},x}^2 = \omega_{\textrm{dc},y}^2 > 0$), as it can be realized in the central horizontal plane of a ring trap with a positive dc bias on the ring electrode. Due to radial symmetry, the ion's equilibrium positions will move in a straight line between the rf and the dc node as the rf power is varied.

When radial symmetry is broken (\eg, $\omega_{\textrm{dc},x}^2 < \omega_{\textrm{dc},y}^2$) the ion's trace changes to a hyperbola (\figref[b]{2d_potential_configs}).
The dashed line indicates the full hyperbola and its asymptotes.
As the rf potential still features rotational symmetry, the hyperbola's asymptotes are parallel to the dc potential's principal axes.

For now we have assumed the rf and dc potentials to be confining in all directions.
In a typical linear Paul trap, the dc potential will be used to create dc confinement in the axial direction.
Since the dc potential has to obey Laplace's equation [$\Delta \phi_{dc}(\vec{r}) \equiv 0$], it must be repulsive in at least one of the radial directions.
\figref[c]{2d_potential_configs} shows the same configuration as \figref[b]{2d_potential_configs}, but with an unstable dc trapping in the $\unitvec{x}$ direction ($\omega_{\textrm{dc},x}^2 < 0$), resembling the potentials in the transverse plane of a linear Paul trap.
The extreme values of the two potentials do not reside on the same branch of the hyperbola but are seperated by its pole.
When decreasing the rf power and such lowering the rf confinement, the ion's equilibrium position will move away from both, the rf minimum and the dc saddle point, finally escaping from the trap when the rf confinement becomes weaker than the dc repulsion.

In the most general case (\figref[d]{2d_potential_configs}) also the rf potential's radial symmetry is broken ($\omega_{\textrm{rf},x}^2 > \omega_{\textrm{rf},y}^2$) and the directions of the dc potential's principal axes are rotated with respect to the rf potential.
This causes a shearing and rotation of the hyperbolic trace, featuring nonperpendicular asymptotes.

A particularly interesting consequence of nondegenerate radial rf trap frequencies is the existence of a nonvanishing axial rf component.
In the quadrupole approximation the rf potenial can be written as
\begin{equation}
	\phi_\textrm{rf}^\textrm{real}(\vec{r},t) = \sum \limits_k \alpha_k(V)\cos(\Omega t) r_k^2,
\end{equation}
where the $\alpha_k$ depend on the actual geometry of the trap and the rf voltage $V$.
Using \eqnref{pseudopotential} and $q_k = \frac{2 Q}{m \Omega^2} \alpha_k$ the rf trap frequencies are given by
\begin{equation}
	\omega_{\rm{rf},k}^2 = \frac{Q^2}{2 m^2 \Omega^2} \alpha_k^2.
\end{equation}
Since the rf potential fullfills Laplace's equation $\Delta \phi_{\textrm{rf}}^{\textrm{real}}(\vec{r},t) \equiv 0 \;\forall t$, it follows that
\begin{eqnarray}
	&&\Delta \phi_{\textrm{rf}}^{\textrm{real}}(\vec{r},t) = 2 \cos(\Omega t) \sum\limits_k \alpha_{k}  \equiv 0 \nonumber \\
	&&\Rightarrow \sum\limits_k \alpha_{k} \equiv 0.
\end{eqnarray}
A vanishing axial rf component $\omega_{\textrm{rf},z}$ requires $\alpha_z = 0$, which immediately yields
the degeneracy of the radial rf trap frequencies,
\begin{equation}
	\alpha_x = - \alpha_y \quad \Rightarrow \quad \omega_{\textrm{rf},x}^2 = \omega_{\textrm{rf},y}^2.
\end{equation} 

\begin{figure}
\centering
\includegraphics[width=0.60\columnwidth]{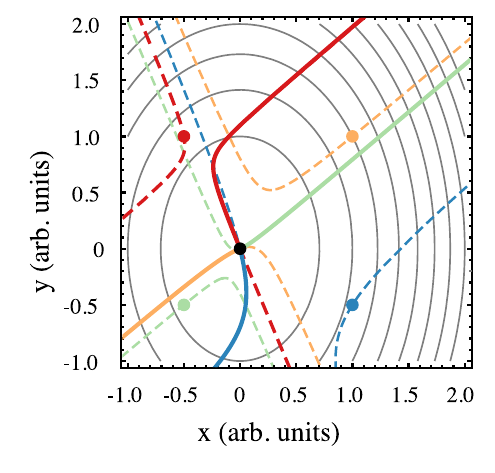}
\caption{(Color online) A combination of different dc potential shifts for a potential configuration as shown in \figref[d]{2d_potential_configs}. Rf (pseudo-)potentials isolines are shown in gray. The ion's traces (thick lines) are shown for the original dc potential (yellow lines) and the shifts of its origin to $(1.0,-0.5)$ (blue lines), $(-0.5,-0.5)$ (green lines) and $(-0.5,1.0)$ (red lines). Dashed lines show the respective full hyperbolas; respectively colored filled circles the dc potentials' origins. All traces converge to the rf node (filled black circle) in the limit of infinite rf power. Near the rf node, the set of all trajectories resembles a star, as expected by intuition. 
The directions of the hyperbola's asymptotes are unchanged by the shifts.}
\label{fig:2d_potential_variation}
\end{figure}

The above considerations allow us to detect excess micromotion by monitoring the ion's position while varying the applied rf power. Using \eqnref{equilibriumEffectiveSimple} (adjusted for a rotation of the dc potential's principal axes), in principle, the measurement of a single trace of the ion's movement during an rf variation without any variation of dc voltages is sufficient to extrapolate the location of the rf node and therefore the location of zero excess micromotion. 

In practice, insufficient knowledge of the trapping parameters (\eg, relative rotation of the rf and dc potential's principal axes, electric field generated by compensation electrodes) might require the combination of variations of the rf power and several dc voltage settings.
An example of such a combination is shown in \figref{2d_potential_variation}. The rf and dc potential configuration is the same as in \figref[d]{2d_potential_configs}, with three additional settings for compensation voltages, shifting the origin of the original dc potential (yellow lines) to $(1.0,-0.5)$ (blue lines), $(-0.5,-0.5)$ (green lines) and $(-0.5,1.0)$ (red lines). The dc potential's origins are marked by the respectively colored dots. The ion's traces (thick lines) for all of these settings converge to the rf node in the limit of infinite rf power. A parallel analysis of the measurements for different dc configurations then yields the position of the rf null as well as the required dc compensation voltages.

\section{Experimental setup}
\label{sec:experimental_setup}

\begin{figure}
\centering
\includegraphics[width=1\columnwidth]{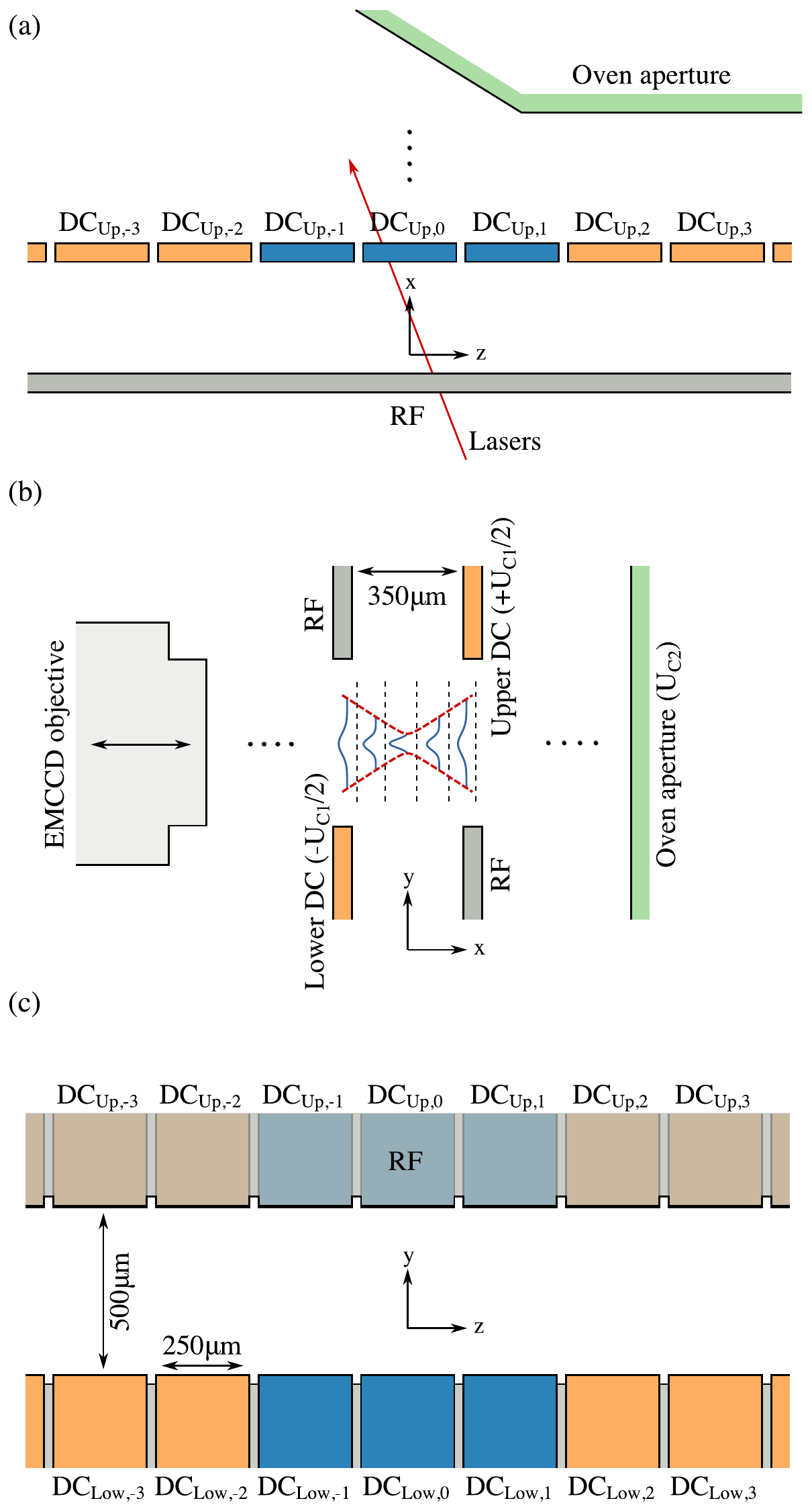}
\caption{(Color online) Experimental setup (not to scale):
(a) Top view (along the $-\unitvec{y}$ direction) including laser beams crossing the trap (enclosing angles of approximately \SI{35}{\degree} with the $\unitvec{x}$ direction) and showing the bend of the oven aperture. The oven aperture is not shown in (c).
(b) Side view (along the $\unitvec{z}$ direction) showing the upper and lower electrode layers (dc and rf) and the Yb oven aperture used as a compensation electrode (voltage U$_{\rm{C2}}$). The second compensation voltage is applied by antisymmetrically biasing all upper and lower dc electrodes (voltage U$_{\rm{C1}}$). Additionally a schematic of the focus-scanning 3D determination of the ion's position [see \secref{FocusScanningPositionDetermination}] is shown. The light gathering objective is mounted on a motorized traveling stage, which is used to sweep the focal plane (vertical dashed black lines) while imaging a series intensity distributions (solid blue lines). The ion's $x$ coordinate is determined from the fit of a Gaussian beam waist to the spot sizes at the different focus settings (thick dashed red lines). The $y$ and $z$ coordinates are taken as the center of a 2D Gaussian fitted to the intensity distribution imaged closest to the $x$ coordinate.;
(c) Frontal view (along the $\unitvec{x}$ direction) of the upper and lower dc electrodes (colored) and  rf electrodes (semitransparent gray).}
\label{fig:trap_setup}
\end{figure}

A schematic of the region of the microstructured segmented linear Paul trap used in the experiments presented here is shown in \figref{trap_setup}.
The trap electrodes have a separation of \SI{500}{\micro\meter} in the $\unitvec{y}$ direction and \SI{350}{\micro\meter} in the $\unitvec{x}$ direction.
The dc electrodes are divided into segments of \SI{250}{\micro\meter} width (labeled dc$_{\textrm{Up},i}$ and dc$_{\textrm{Low},i}$ in \figref{trap_setup}).
The full trap features 33 pairs of dc electrodes composing a wide section (described above) and a narrow section (electrode width, \SI{100}{\micro\meter}; $\unitvec{y}$ separation, \SI{250}{\micro\meter}) that are connected by a tapered transfer section.  The trap and the supporting experimental setup have been described in detail in \cite{Schulz2008, Kaufmann2012, Baig2013}.

For this work we trapped single \ybgion{} ions, which are produced by a two-photon ionization process from neutral ytterbium vapor emitted from the atom oven \cite{Johanning2011}.
The ion is Doppler cooled at the $S_{1/2}-P_{1/2}$ transition.
The ionization laser (\SI{399}{\nano\meter}), the cooling laser (\SI{369}{\nano\meter}), and two additional repumpers (\SI{638}{\nano\meter} and \SI{935}{\nano\meter}) are overlapped outside the trap \cite{Kaufmann2012} and propagate through the center of the trapping region, enclosing angles of approximately \SI{35}{\degree} with the $\unitvec{x}$ direction and \SI{89}{\degree} with the $\unitvec{y}$ direction (see \figref[a]{trap_setup}).

For basic trapping the central electrode pair (dc$_{\textrm{Up/Low},0}$) is set to \SI{-1}{\volt}, providing axial confinement, whereas all other dc electrodes are set to ground.
The rf electrodes are driven through a helical resonator with a voltage of \SI{210}{\volt_{\rm{pp}}} at a frequency of \SI{13.2}{\mega\hertz}. The secular frequencies of an ion trapped with this configuration are $\omega_{x^{\prime}}$ = \SI[product-units = single]{2\pi x 1.054}{\mega\hertz}, $\omega_{y^{\prime}}$ = \SI[product-units = single]{2\pi x 1.112}{\mega\hertz} and $\omega_z$ = \SI[product-units = single]{2\pi x 199}{\kilo\hertz}.
The principal radial axes of the trapping potential ($\unitvec{x}^\prime$ and $\unitvec{y}^\prime$) are rotated by about \SI{64}{\degree} with respect to the $\unitvec{x}$ and $\unitvec{y}$ axes.

Static and approximately uniform fields for micromotion minimization are created using three compensation voltages U$_{\textrm{C}}$ (see \figref[b]{trap_setup}).
U$_{\textrm{C1}}$ is added as an antisymmetrical bias voltage to all upper (+U$_{\textrm{C1}}/2$) and lower (-U$_{\textrm{C1}}/2$) dc segments, causing a field roughly along the $\unitvec{x}+\unitvec{y}$ direction near the trap axis.
A second compensation voltage, U$_{\textrm{C2}}$, is applied to the aperture of the nearby atom oven.
The aperture extends mainly in the $y$-$z$ plane and its size largely exceeds the dimensions of the central trap slit, creating a field mainly along the $\unitvec{x}$ direction.
A bend of the aperture along the $\unitvec{y}$ direction near the trapping site causes a non-negligible $z$ component of the compensation field created by U$_{\textrm{C2}}$ (see \figref[a]{trap_setup}). Therefore a counteracting field is created by adding a calibrated bias voltage U$_\textrm{B}(\rm U_\textrm{C2})$ antisymmetrically to all outer left (dc$_{\textrm{Up/Low},i}$, $i \leq -2$) and outer right ($i \geq 2$) dc electrodes (colored orange), which cancels the unwanted $z$ component of the aperture's field.
The third compensation voltage U$_{\textrm{C3}}$ is applied antisymmetrically to the same outer electrodes if control of the ion's axial position is desired.

The ion is detected by imaging fluorescence of the cooling transition along the $-\unitvec{x}$ direction on an EMCCD (Andor iXon X3 Blue) (see \figref[b]{trap_setup}): 
The light gathering objective (numerical aperture of \num{0.4}) \cite{Schneider2007} is mounted on a motorized traveling stage and allows a positioning of the focus of the imaging system with a resolution of \SI{8}{\nano\meter} along the $x$ axis.
The objective collects the ion's fluorescence with a magnification of approximately \num{12.5} onto the chip of the EMCCD, consisting of 512 $\times$ 512 pixels with an edge length of \SI[product-units = single]{16}{\micro\meter} $\times$ \SI[product-units = single]{16}{\micro\meter}.

\section{Method and Results}
\label{sec:MethodAndResults}

\subsection{Focus-scanning position determination}
\label{sec:FocusScanningPositionDetermination}

\begin{figure}
\centering
\includegraphics[width=0.85\columnwidth]{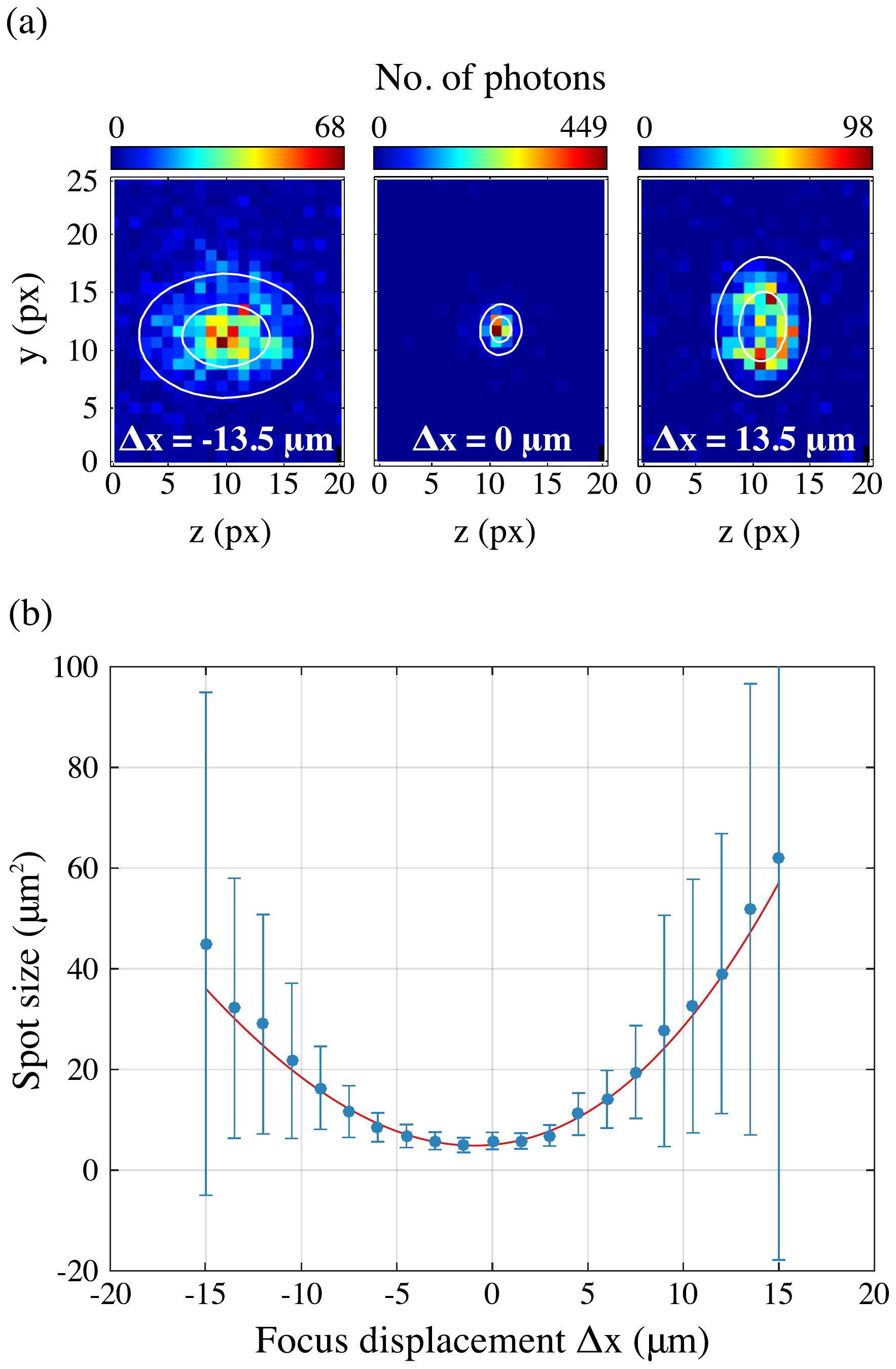}
\caption{(Color online) Focus-scanning 3D determination of the ion's position. (a) Fitting t2D Gaussians to the intensity distributions taken at different focus displacements yields the center coordinates ($y$,$z$) of the distribution as well as the spot size. Exemplary fits are given for focus displacements of \SI{\pm13.5}{\micro\meter} and \SI{0}{\micro\meter}. White lines indicate the distributions isolines at one and two FWHM. (b) The ion's $x$ coordinate is deduced by fitting the spot size of a Gaussian beam waist to the spot sizes obtained from the single images. Error bars for large focus displacements are overestimated as the assumption of a gaussian beam only holds near to the focus. The center coordinates of the intensity distribution imaged with the focus setting closest to the ion's $x$ coordinate are taken as the ion's $y$ and $z$ coordinates.}
\label{fig:x_position_fit}
\end{figure}

Since our experiment features imaging of the ion only from a single direction, we determine the full 3D position of the ion in our trap during the rf and dc variation by employing focus-scanning imaging (see \figref[b]{trap_setup}).
For every position determination, we take several images of the ion with the focus swept over \SI{\pm15}{\micro\meter}. 
For every single image a 2D Gaussian is fitted to the intensity distribution, yielding $(y, z)$ center coordinates of the distribution as well as the Gaussian spot size.
The $x$ coordinate of the ion is then deduced by fitting the spot size of a Gaussian beam to the spot sizes obtained from the single images (see \figref{x_position_fit}).
Then the center coordinates of the intensity distribution imaged with the focus setting closest to the ion's $x$ coordinate are taken as the $y$ and $z$ coordinates of the ion.
Using this procedure, one is able to determine the average ion position with a precision that is significantly below the wavelength of the fluorescence light.
In principle, the precision scales as the inverse square root of the observed fluorescence and is limited only by its signal to noise ratio and the mechanical stability of the imaging system. 
For an average number of about 7000 photons per image (observed scattering rates of about \SIrange{200}{400}{\kilo\hertz}) and images taken at 25 focal positions, we routinely achieve a precision of $(\sigma_x,\sigma_y,\sigma_z) = (\SI{98}{\nano\meter}, \SI{39}{\nano\meter}, \SI{35}{\nano\meter})$.

\subsection{Micromotion minimization in two dimensions}
\label{sec:MicromotionMinimizationIn2d}

To minimize micromotion in the radial plane at a specific axial position we choose a set of voltage settings $\{(U_{\textrm{C1}}, U_{\textrm{C2}})_{i}\}$ and apply a set of rf powers $\{P_{\textrm{rf},{j}}\}$.
For all combinations $(i,j)$ a focus-scanning position determination is carried out, yielding the ion's equilibrium positions $\{\vec{r}_{\textrm{0},i,j}\}$.
To suppress possible systematic effects of a quasicontinuous change in the rf power, we iterate over the different rf powers not in ascending or descending order, but over some permutation of the chosen set.

We then model our effective trapping potential as a superposition of the rf pseudopotential, the unshifted dc potential, and two additional linear potentials as created by our compensation electrodes,
\begin{eqnarray}
	\label{eq:fitmodel}
	\Phi_\textrm{eff}^{(i,j)}(x,y) &=& \frac{m}{2} \frac{P_{\textrm{rf},j}}{P_{\textrm{0}}} (\vec{r} - \vec{r}_{\textrm{rf}})^T \matr{R}_{\alpha_\textrm{rf}} \matr{M}_{\omega_\textrm{rf}^2} \matr{R}^{-1}_{\alpha_\textrm{rf}} (\vec{r} - \vec{r}_{\textrm{rf}}) \nonumber\\
	&& + \frac{m}{2} (\vec{r} - \vec{r}_{\textrm{rf}})^T \matr{R}_{\alpha_\textrm{dc}} \matr{M}_{\omega_\textrm{dc}^2} \matr{R}^{-1}_{\alpha_\textrm{dc}} (\vec{r} - \vec{r}_{\textrm{rf}}) \nonumber\\
	&& + s_\textrm{C1} (U_\textrm{C1,i} - \tilde{U}_\textrm{C1}) \unitvec{y}^T \matr{R}_{\alpha_\textrm{C1}} (\vec{r} - \vec{r}_{\textrm{rf}}) \nonumber\\
	&& + s_\textrm{C2} (U_\textrm{C2,i} - \tilde{U}_\textrm{C2}) \unitvec{x}^T \matr{R}_{\alpha_\textrm{C2}} (\vec{r} - \vec{r}_{\textrm{rf}}),
\end{eqnarray}
where $\matr{R}_{\alpha}$ is a rotation by $\alpha$, $\matr{M}_{\omega_\textrm{rf}^2}$ and $\matr{M}_{\omega_\textrm{dc}^2}$ are the diagonal matrices of the squared rf and dc trap frequencies ($\omega_{{\textrm{rf}_\textrm{0}},k}^2$ and $\omega_{\textrm{dc},k}^2$), and $s_\textrm{C1}$ and $s_\textrm{C2}$ are geometrical factors depending on the compensation electrode shape and placement.
$\tilde{U}_\textrm{C1}$ and $\tilde{U}_\textrm{C2}$ are the compensation voltages that cancel the stray potentials and shift the origin of the dc potential to the position of the rf null ($\vec{r}_\textrm{rf}$).
The trap frequencies are determined from independent measurements by excitation of the ion's motion with a voltage modulation of one dc segment for various settings of the applied rf power $P_\textrm{rf}$.
All remaining parameters are determined by fitting the equilibrium positions obtained from \eqnref{fitmodel} to the $\{\vec{r}_{\textrm{0},i,j}\}$.

\begin{figure}
\centering
\includegraphics[width=\columnwidth]{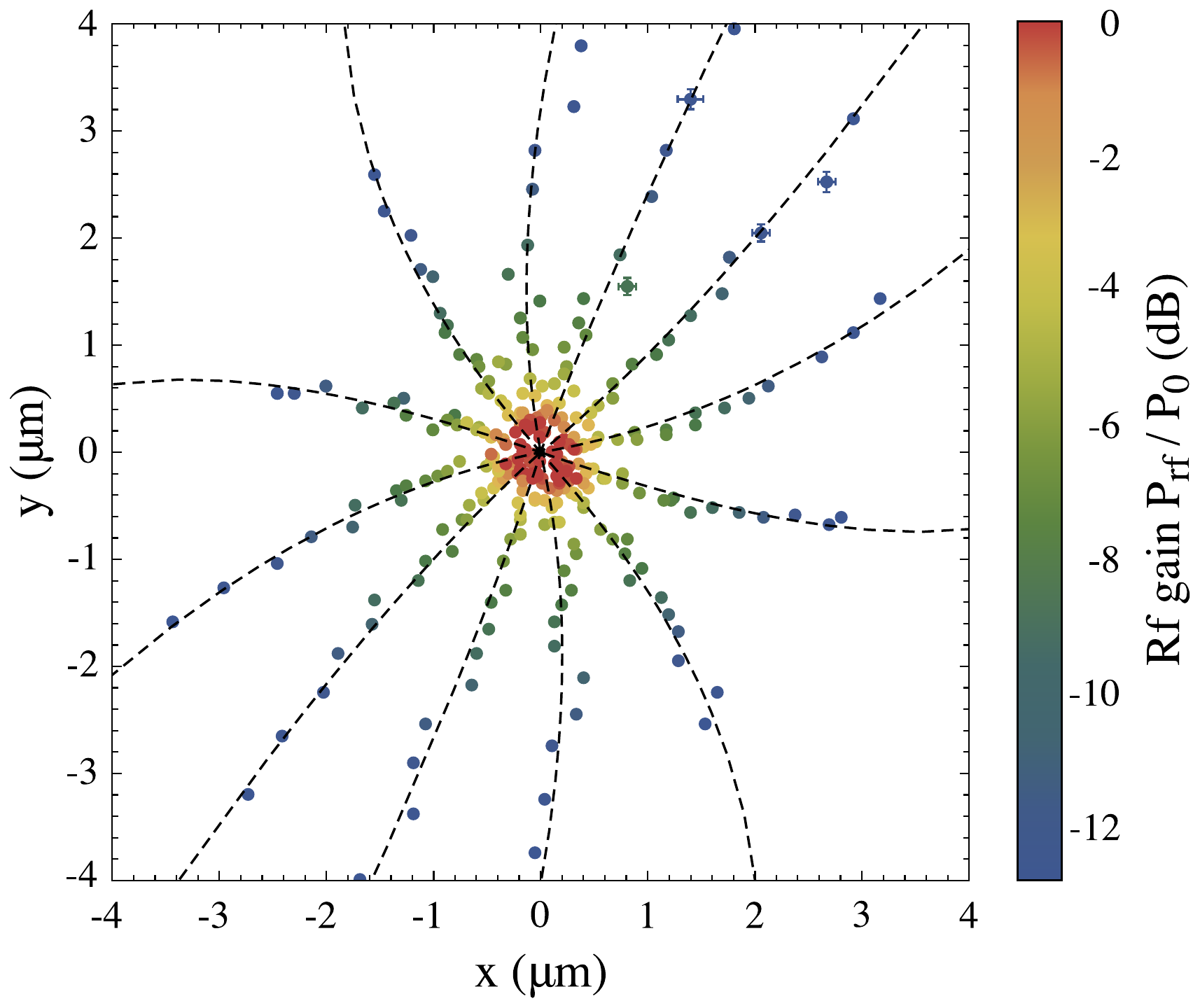}
\caption{(Color online) A 2D micromotion minimization performed in the radial plane perpendicular to the trap axis. The $x$ axis is along the line of sight of our imaging optics. 
	Optimization was carried out using 12 settings of compensation voltages (U$_\textrm{C1}$, U$_\textrm{C2}$) and 25 rf power settings.
	Colored filled circles represent the ion's positions with their respective rf powers, and example error bars indicating $\pm\sigma_x$ and $\pm 2 \sigma_y$ are given for four points in the upper-right quadrant. Fit results of the ion's trajectories are indicated by dashed lines and the rf node is marked by the filled balck circle.
	The residual electric stray field uncertainty along the principal radial axes after the optimization is determined to $(\Delta\varepsilon_{\textrm{stray},x^{\prime}}, \Delta\varepsilon_{\textrm{stray},y^{\prime}}) = (\SI{0.09}{\volt\per\meter}, \SI{1.09}{\volt\per\meter})$ at respective trap frequencies of $\omega_{x^{\prime}} = \SI[product-units = single]{2\pi x 132.5}{\kilo\hertz}$ and $\omega_{y^{\prime}} = \SI[product-units = single]{2\pi x 387.0}{\kilo\hertz}$, where the principal axes of the trapping potential ($\unitvec{x}^\prime$, $\unitvec{y}^\prime$) are rotated by  \SI{64.8}{\degree} counterclockwise with respect to the $\unitvec{x}$ and $\unitvec{y}$ axes.
	}\label{fig:compensation2d}
\end{figure}

\figref{compensation2d} shows the results of an optimization procedure with \num{12} settings for the compensation voltages $(U_{\textrm{C1}}, U_{\textrm{C2}})$ and variation of the applied rf power over \SI{12.5}{\decibel} in \num{25} steps (varying the rf voltage from \SI{420}{\volt_{\rm{pp}}} to \SI{99}{\volt_{\rm{pp}}}). The circles represent the ion's positions $\vec{r}_{\textrm{0},i,j}$ with color-coded rf power, the dashed line represents the fit, and the position of the rf node $\vec{r}_\textrm{rf}$ is marked by the central black circle.
After applying the optimized compensation voltages as yielded by the fit, we perform a similar measurement but without any variation of the compensation voltages. While lowering the rf confinement, we detect no change in the ion's average position before it escapes the trap (at an rf voltage of about \SI{62}{\volt_{\rm{pp}}}), where the escape happens more rapidly than the time resolution of our position determination.
Using \eqnref{equilibriumEffectiveSimple} and the weighted standard deviation of the ion positions along the principal radial axes of our trapping potential, we derive a residual electric stray field uncertainty of $(\Delta\varepsilon_{\textrm{stray},x^{\prime}}, \Delta\varepsilon_{\textrm{stray},y^{\prime}}) = (\SI{0.09}{\volt\per\meter}, \SI{1.09}{\volt\per\meter})$ at respective trap frequencies of $\omega_{x^{\prime}} = \SI[product-units = single]{2\pi x 132.5}{\kilo\hertz}$ and $\omega_{y^{\prime}} = \SI[product-units = single]{2\pi x 387.0}{\kilo\hertz}$, where the principal axes of the trapping potential ($\unitvec{x}^\prime$, $\unitvec{y}^\prime$) are rotated by  \SI{64.8}{\degree} counterclockwise with respect to the $\unitvec{x}$ and $\unitvec{y}$ axes.

\subsubsection*{Rf power dependence of the trap structure}
\label{sec:TrapHeating}
During the measurements the rf drive power is varied, so the energy deposition at the trap structure will be changed.
The accompanying change in temperature might affect the trap geometry due to thermal expansion and might result in an rf power dependence of the rf node.
The rf node yielded by the trajectory analysis might then be different from the rf node present during normal trap operation. 

If the time between changes in the rf power is short compared to the time constant for temperature changes in the trap structure, the rapid changes in the rf drive will average to an effective rf power throughout the measurement.
Thus the temperature of the system does not change significantly if the effective power during the measurement is matched to the constant rf power during normal operation.
This matching can always be achieved by the choice of rf powers that are sampled during the minimization process.

In our particular setup the trap is composed of gold-plated aluminium oxide glued and wire-bonded to an actively cooled aluminium oxide holder \cite{Kaufmann2012}. Given the low thermal resistivity of our trap-holder setup and its good thermal coupling to the cooling system, we estimate the time constant for temperature changes of the trap-holder system to be large compared to the time between changes in the rf power for our measurements (a few tens of milliseconds).
Also, our comparatively low maximal rf trapping voltage (\SI{420}{\volt_{\rm{pp}}}) (and, thus, low heating power) limits the system to only small temperature changes.
In addition, the order in which measurements at different rf powers are carried out adds to the averaging effect. Thus, the conditions for an effective constant rf drive power are fulfilled in the experiments presented here.

A non-negligible change in the trap geometry caused by the changes in the rf power would manifest itself in a discrepancy between model and measurement data.
A change in the trap geometry would yield both a change in the rf trap frequencies $\omega_{\textrm{rf}, k}$ [\eqnref{pseudopotential}] not compatible with the changes predicted by the model and a change in the dc trap frequencies $\omega_{\textrm{dc}, k}$ [\eqnref{dcQuadrupole}] that are assumed constant. These effects would alter the ion trajectory as a function of the rf power [\eqnref{equilibriumEffectiveSimple}], which is not observed in the experiment (\figref{compensation2d}).

In general, even if a trap structure exhibits a time constant for rf-drive-induced temperature changes that is comparable to the time required to take a single image of the ion, and if thermal deformation of the trap geometry is of concern, the latter can be avoided.
The single-image exposure can always be shortened such that the averaging condition mentioned above is fulfilled.
The whole measurement sequence will then be executed repeatedly and the single-image position determination is performed by averaging over repetitions of the measurement sequence, with only a minor effect of increased readout noise.

If constant rf power is mandatory, the method presented here can also be carried out by a variation of a dc voltage bias applied to the rf or dc electrodes instead of the rf power variation \cite{Schneider2005}.
Under variation of a dc bias symmetrically applied to the rf or dc electrodes, the rf trap frequencies will remain constant, while the dc trap frequencies change.
As in the case of an rf power variation the ion's equilibrium position will follow hyperbolic trajectories [\eqnref{equilibriumEffectiveSimple}].

\subsection{Micromotion minimization in three dimensions}
\label{sec:MicromotionMinimizationIn3d}

By construction, an ideal linear trap features no axial rf electric field component and therefore no axial micromotion.
For all realizations with finite-size electrodes, particularly for segmented traps, regions of non-negligible axial micromotion cannot be avoided, although recent results have demonstrated trap designs with small residual axial rf field components \cite{Narayanan2011,Doret2012,Wilpers2012,Pyka2014}.

The focus of segmented traps, such as the one used here, often lies in the storage and manipulation of ion crystals. Such operations, \eg, shuttling, splitting, and merging, make use of an extended region along the almost-micromotion-free symmetry axis of the linear trap.
In these cases a space curve of positions of minimized radial micromotion along the axial direction may be used to optimize the operation of the trap.
In a similar sense, when experimenting with extended linear ion crystals oriented along this axis, finding a single point of vanishing axial micromotion is not meaningful. Any axial offset from this ideal position, as long as it is much smaller than the extent of the crystal will have a small effect. So the level of precision aimed for in the determination of the axial position will be of the order of a micrometer or slightly less.

\begin{figure}
\centering
\includegraphics[width=\columnwidth]{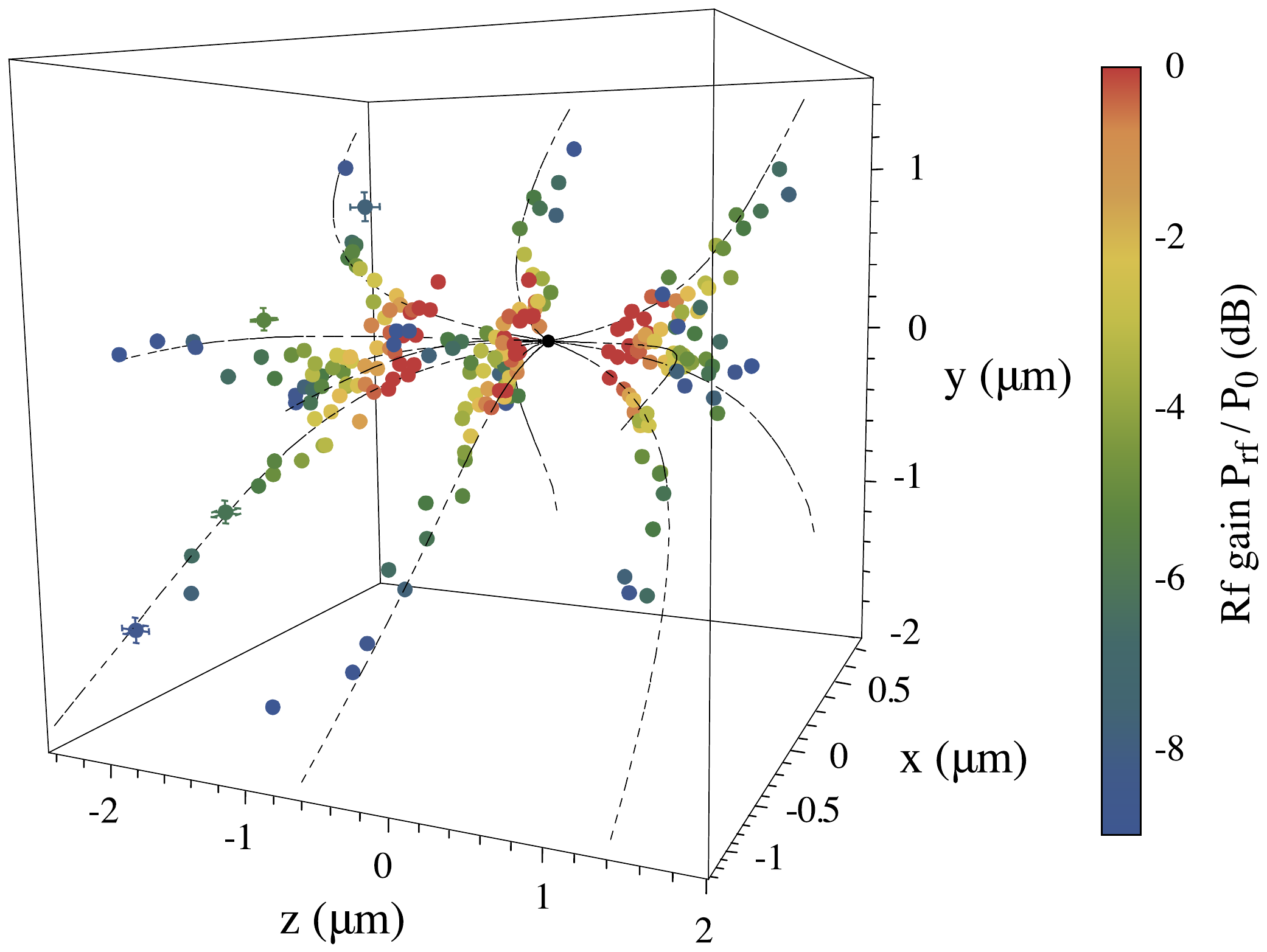}
\caption{(Color online) Results of a 3D optimization procedure performed in a trap region with non-negligible axial rf field strength. Minimization was performed using 4 (U$_\textrm{C1}$,U$_\textrm{C2}$) $\times$ 3 (U$_\textrm{C3}$) settings of compensation voltages. Filled circles represent the average ion positions obtained from focus-scanning imaging, with their colors indicating the respective rf powers. Example error bars indicating $\pm\sigma_x$ and $\pm 2 \sigma_{y,z}$ are given for four points. A fit of the trajectories is indicated by dashed lines. The filled black circle shows the position of the rf node.}
\label{fig:3d_data}
\end{figure}

If the axial positions of trapped ions are not constrained by other experimental requirements, minimization of axial as well as radial micromotion can be performed. 
A 3D micromotion minimization is demonstrated qualitatively by extending the procedure for 2D minimizations by a variation of the axial compensation voltage $U_{\textrm{C3}}$.
The model of the effective trapping potential given in \eqnref{fitmodel} is complemented by adding the respective terms for axial components of the rf pseudopotential, dc potential, and linear potential created by $U_{\textrm{C3}}$.
\figref{3d_data} demonstrates a 3D optimization procedure performed in a region of our trap suffering from significant axial rf electric field strength. The minimization was performed with 4 different settings for the radial compensation voltages $(U_{\textrm{C1}},U_{\textrm{C2}})$ and 3 different settings for the axial compensation voltage $U_{\textrm{C3}}$, yielding 12 ion trajectories of varying rf powers. Filled circles represent the average ion positions, with their colors indicating the respective rf powers. A fit of the trajectories from the 3D effective trapping potential are given by dashed lines. As described in \secref{Theory} all trajectories approach the rf node (filled black circle) for increasing rf field strength.

\section{Performance}

In \cite{Haerter2013} a comparison of several micromotion minimization approaches is given in terms of various figures of merit, such as the residual electric stray field uncertainty, the micromotion amplitude, and the average kinetic energy of micromotion, which relates to the second-order Doppler shift.
For a specific experiment, one of these numbers might be more suitable or relevant than the others.  The micromotion amplitude and the average kinetic energy depend, to different extents, on the ion mass and on the drive and secular frequencies. These numbers vary substantially throughout the ion trapping community, therefore the residual electric stray field uncertainty $\Delta\vec{\varepsilon}$, from which the other quantities mentioned above can be derived, is the relevant quantity that can easily be compared among experiments.
In general, by increasing the rf drive power, one can, within the constraints given by the experiment, reduce the effect of micromotion. 

In \tabref{performanceComparison} the outcomes of several minimization methods are listed.
The final level of precision of micromotion minimization can be scaled in each method by changing the averaging time for data taking, as long as the stray fields to be compensated are stable. 
Ultimately, as our minimization approach relies on imaging, the position determination and, in turn, the determination of the rf null location are shot noise limited.  Thus another question to ask is the sensitivity of a minimization method, that is, on what timescale what level of precision is obtained. Here, almost no information was found in the literature. 

The observed fluorescence rate is limited by the $S_{1/2}-P_{1/2}$ transition of \ybgion and its linewidth $\Gamma=\SI[product-units = single]{2\pi x 19.2}{\mega\hertz}$, the solid angle covered by our light gathering optics (\SI{4}{\percent} of $4\pi$) and its transmission, and the quantum efficiency of our camera (\SI{55}{\percent}). With these constraints and the time required for the translation of our optics (about \SI{400}{\second}), recording 
a data set such as shown in \figref{compensation2d} takes about \SI{620}{\second} and directly yields all compensation voltages. We achieve a sensitivity of the residual electric field in the radial directions of
$(s_{\varepsilon_{\textrm{stray}},x^{\prime}}, s_{\varepsilon_{\textrm{stray}},y^{\prime}}) = (\SI{2.3}{\volt\per\meter\sqrt{\hertz}}, \SI{27.1}{\volt\per\meter\sqrt{\hertz}})$.

\begin{table*}
\caption{Comparison of micromotion minimization methods in terms of the residual electric stray field uncertainty $\Delta\vec{\varepsilon}$. The first six rows of data are an extension of the values given in \cite{Haerter2013}; the last row states the results obtained in this work. If different residual electric stray fields for different trapping directions are reported, only the lowest value per reference is given.}
\begin{ruledtabular}
\begin{tabular}{lcl}
Ref. & Method & $\Delta \varepsilon$\,(V/m)\\
\hline
\cite{Pyka2014} & Photon-correlation spectroscopy & 0.9\\
\cite{Chou2010} / \cite{Akerman2012} / \cite{Chwalla2009} & Micromotional sideband spectroscopy & 7 / 1 / 0.4\\
\cite{Chuah2013} & Ion-cavity emmission spectroscopy & 1.8\\
\cite{Tanaka2012} / \cite{Narayanan2011} & Parametric excitation of secular motion & 6 / 0.4\\
\cite{Haerter2013} & Neutral atom loss & 0.02\\
\cite{Schneider2005} & Monitor displacement & $\leq$ 11.8\\
This work& Trajectory analysis & 0.09\\
\end{tabular}
\end{ruledtabular}
\label{tab:performanceComparison}
\end{table*}

\section{Position determination application: light pressure measurement}
\label{sec:LightPressureMeasurement}

Another application of precise position measurements is the determination of small forces, where trapped ions have been proven to be excellent probes \cite{Knuenz2010, Biercuk2010}. We demonstrate this with the example of the light pressure force \cite{Frisch1933} acting on a laser-cooled ion in a Paul trap.

When micromotion is minimized using any scheme that utilizes a signal from an incoherent scattering process, the light pressure force due to the fluorescence-inducing laser is compensated automatically.
As a consequence, the settings obtained from the minimization are only valid at the scattering rate used
during the process and will be different if the interaction is changed (\eg, change in laser intensity or use of a different transition). This is especially true for coherent manipulation of the ion, as there is no light pressure force present.

If minimized micromotion is required for situations without photon scattering, it is neccessary to perform the minimization for several scattering rates and extrapolate the minimization settings to the zero-scattering level. Given sufficient knowledge about the dc potential, one can monitor the ion's position as a function of the scattering rates and counteract solely the light-pressure-induced shift.

The force corresponding to the time-averaged momentum transfer of absorbed laser photons (or light pressure force) $\vec{F}_\textrm{lp}$ is given by 
\begin{equation}
	\label{eq:lightPressureForce}
	\vec{F}_\textrm{lp}=\mean{\dot{\vec{p_\textrm{ph}}}}_t = \Gamma \hbar \vec{\kappa}
\end{equation}
with rate of absorption $\Gamma$, Planck constant $\hbar$, and laser wave vector $\vec{\kappa}$.

This force can effectively be treated as originating from a stray potential and adds as $F_{\textrm{lp},k}/m$ to the right-hand-side of \eqnref{mathieuEquations}. In analogy with \eqnref{averageDisplacement}, the light pressure force shifts the ion's average position by
\begin{equation}
	\label{eq:lightPressureShift}
	\vec{r}_\textrm{lp} = \frac{1}{m}\matr{M}^{-1}_{\omega^2}  \vec{F}_\textrm{lp}\, ,
\end{equation}
with $\matr{M}^{-1}_{\omega^2}$ being the inverse matrix of the squared secular frequencies. In the frame of the potential's principal axes $\matr{M}^{-1}_{\omega^2}$ is diagonal and \eqnref{lightPressureShift} reduces to $\vec{r}_{lp} = \frac{1}{m} \sum\limits_{k} \frac{F_{\textrm{lp},k}}{\omega_k^2}\unitvec{k}^\prime$ [see \eqnref{averageDisplacement}].

When the light field interacting with the cooling transition is detuned by $\delta$ from resonance, the scattering rate $\Gamma$ and hence the rate of absorption of the ion are given by
\begin{equation}
	\Gamma = \frac{s_0}{1+s_0}\frac{\gamma/2}{1 +(2\delta/\gamma')^2}
\end{equation}
with the on-resonance saturation parameter $s_0 = 2\Omega^2 / \gamma^2$, the Rabi frequency $\Omega$ of the interaction, the natural line width of the transition $\gamma$, and the saturation broadened line width $\gamma' = \gamma\sqrt{1+s_0}$. 

\begin{figure}
\centering
\includegraphics[width=\columnwidth]{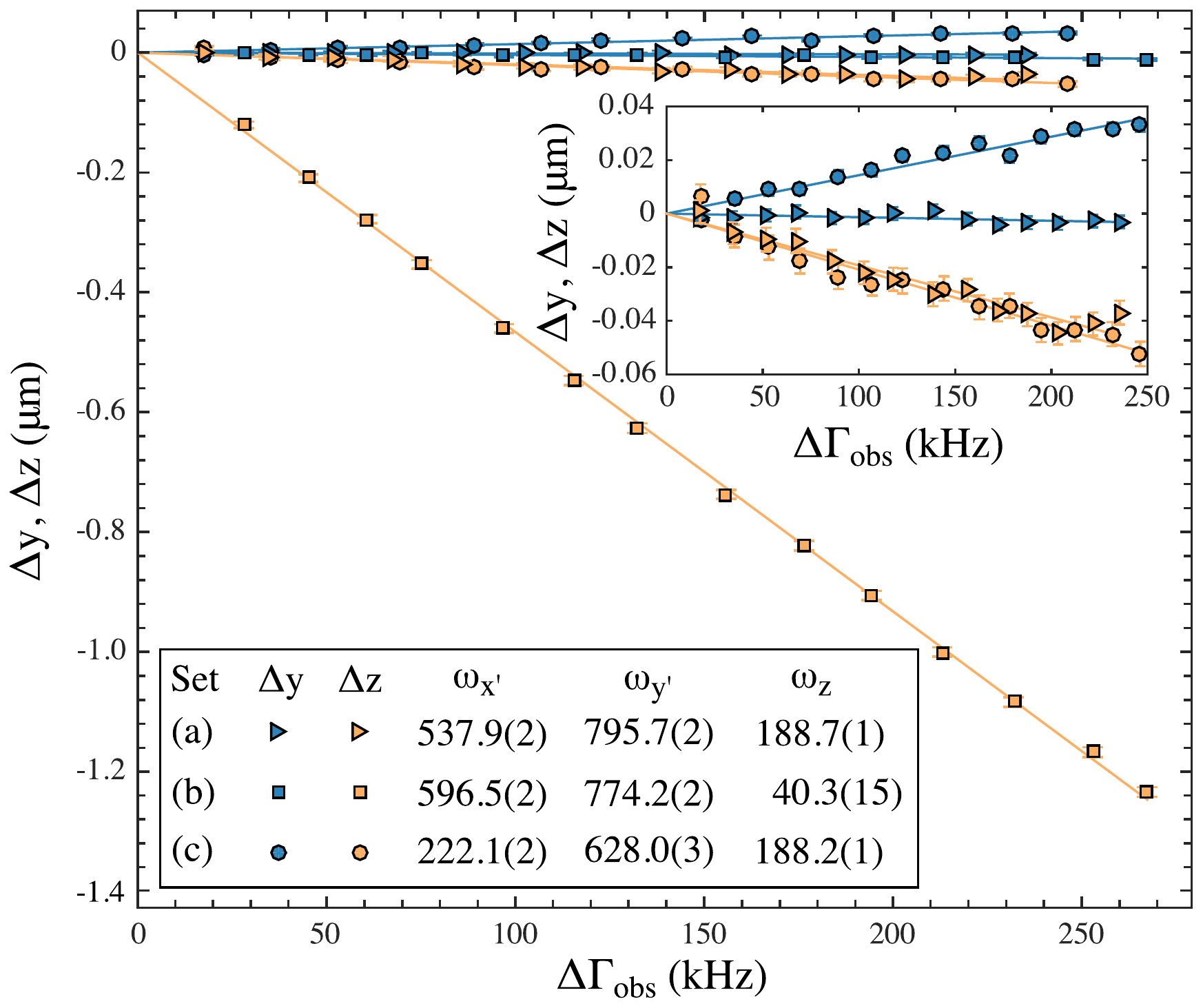}
\caption{(Color online) Shifts of the ion's $y$ (blue symbols) and $z$ (yellow symbols) position as a function of the change in observed scattering rate for three trap configurations. Corresponding symbols and trap frequencies (in \si{\kilo\hertz}) are specified in the legend. Set (a) gives strong confinement in all three directions; set (b) features weakened axial confinement resulting in an increased shift along the $\unitvec{z}$ direction. Set (c) is weakened in terms of radial confinement and causes a substantial shift along the $\unitvec{y}$ direction. Inset: Zoom-in on the region of small shifts. Straight lines are fits of the linear dependence of the position shift on the change of the light pressure force (see text).}
\label{fig:lightPressure}
\end{figure}

The expected light pressure force of a laser saturating ($s_0 = 1$) the $S_{1/2}-P_{1/2}$ transition of \ybgion{} red detuned by $\gamma/2$ is $|\vec{F}_\textrm{lp}| \approx \SI{35}{\zepto\newton}$. 
For the parameters given as setting (a) in Fig.~\ref{fig:lightPressure}, this force would cause a shift of $\vec{r}_\textrm{lp} \approx (\SI{4}{\nano\meter},\SI{1}{\nano\meter},\SI{-50.0}{\nano\meter})$, which is of the order of the level of precision of the position determination reported above for the $z$ component and significantly below for the $x$ and $y$ components. 
To be able to observe the light pressure shift, averaging over several position determinations is a neccessity in our present setup. 
Furthermore, the mechanical stability of our imaging system requires differential measurements to cancel slow movements of the objective mount causing a virtual shift of the ion's position. 

We measure the shift of the ion's center coordinates in the $z$-$y$ plane as a function of the change in the observed scattering rate for different sets of trap frequencies.
The ion's scattering rate is varied by tuning the intensity of the \SI{369}{nm} light field using an acusto-optic modulator in 15 steps.
Exposure times are adjusted such that the amount of collected photons for each intensity setting is approximately constant to avoid systematic effects in the position determination.
We obtain the shifts of the ion's central positions $(\Delta y, \Delta z)$ with respect to the lowest intensity setting and average these shifts over 500 repetitions of the intensity variation for each trapping configuration, which improves the precision to the nanometer
range.
The change in observed scattering rate $\Delta\Gamma_\textrm{obs}$ is determined by extracting the number of collected fluorescence photons from the ion images and referencing to the lowest intensity setting as well.
Using a part of the image with negligible contributions from the ion's intensity distribution we monitor and correct for the background.

\figref{lightPressure} shows the shifts of the ion's average position in $\unitvec{y}$ (blue symbols) and $\unitvec{z}$ direction (yellow symbols) versus the change in observed scattering rate for three sets of trap frequencies: (a) strong confinement in all three directions, (b) weakened axial confinement, and (c) weakened radial confinement. Detailed trap frequencies and corresponding symbols are given in the legend in \figref{lightPressure}.
The inset shows a zoom-in where data from the weak axial confinement have been left out for clarity.
The linear dependence of the ion's position shift on the observed scattering rate and thus the light pressure force given by \eqnref{lightPressureShift} and \eqnref{lightPressureForce} is clearly visible.

We fit a linear model, $\Delta r_{\textrm{lp},k} = \alpha_{k} \cdot \Delta\Gamma_\textrm{obs} + \beta_{k} \, (k=y,z)$, to the data, where $\alpha_{k} = \gamma \frac{\hbar}{m} \left(\matr{M}^{-1}_{\omega^2} \vec{\kappa}\right)_{k}$ and $\gamma$ is a scaling factor accounting for the overall efficiency of our imaging system ($\Gamma = \gamma \cdot \Gamma_\textrm{obs}$).
When we weaken the axial confinement of the ion but keep a comparable radial confinement [going from case (a) to (b)], the movement in $\unitvec{z}$ direction increases from $\alpha_{z}^{\rm(a)} = \SI{-0.19(3)}{\nano\meter\per\kilo\hertz}$ to $\alpha_{z}^{\rm(b)} = \SI{-4.66(4)}{\nano\meter\per\kilo\hertz}$ whereas the shift in the $\unitvec{y}$ direction ($\alpha_{y}^{\rm(a)} = \SI{-0.01(1)}{\nano\meter\per\kilo\hertz}$) changes only negligibly.
Keeping the axial confinement but lowering the radial confinement [going from case (a) to (c)], we observe no change in the shift along the $\unitvec{z}$ direction but an increased shift along $\unitvec{y}$ of $\alpha_{y}^{\rm(c)} = \SI{0.144(13)}{\nano\meter\per\kilo\hertz}$, which agrees with the approximate orientation of the trapping potential's principal axes.

Since $\matr{M}^{-1}_{\omega^2}$ is not diagonal in our frame of observation, the ion's displacements cannot in general be associated with only one of the trap frequencies, and as the principal axes of the dc and rf effective trapping potentials differ, changing the radial confinement also changes the orientation of the radial trapping axes.
For the shifts of the ion position along the $\unitvec{z}$ direction, however, $\alpha_z = \gamma \frac{\hbar}{m} \frac{\kappa_z}{\omega_z^2}$, as the axis of our linear trap is aligned in parallel with $\unitvec{z}$. 
So the ratio of $\alpha_z$ for two different axial confinements should be equal to the squared inverse ratio of the trap frequencies $\omega_z$. Ratios obtained from the measurements displayed in \figref{lightPressure} and the respective trap frequency ratios for the axial position shifts are listed in \tabref{lightPressureRatios}, yielding a good agreement.

\begin{table}
	\caption{Comparison of the ratios of $\alpha_z$ as obtained from our fits and the squared inverse ratios of the trap frequencies $\omega_z$ ($\alpha_z \propto \frac{1}{\omega_z^2}$) for three trapping configurations (a), (b) and (c) (see text and \figref{lightPressure}).} 
\begin{ruledtabular}\begin{tabular}{@{}c|lll}
($i$),($j$) & \multicolumn{1}{c}{(a),(b)} & \multicolumn{1}{c}{(c),(b)} & \multicolumn{1}{c}{(c),(a)} \\
\hline
$\left(\omega_{z}^{(i)} / \omega_{z}^{(j)}\right)^2$ & 0.046(3) & 0.046(3) & 0.9947(15)\\[2mm]
$\alpha_z^{(j)}/\alpha_z^{(i)}$ & 0.041(5) & 0.044(5) & 0.92(16)\\
\end{tabular}
\end{ruledtabular}
\label{tab:lightPressureRatios}
\end{table}

The best sensitivity and uncertainty are achieved for lightweight particles and low trap frequencies along the direction of the scattering force [see \eqnref{lightPressureShift}].
As the sensitivity depends on position accuracy, the measurement can again be optimized for either precision or speed.
For our setup we report a sensitivity for the measurement of the light pressure force along the $\unitvec{z}$ direction of $s_{{\textrm{F}_\textrm{lp}}} = \SI{633}{\yocto\newton\per\sqrt{\hertz}}$ (total exposure of a differential measurement: \SI{38}{\milli\second}), with our minimal absolute uncertainty being  
$\sigma_{{\textrm{F}_\textrm{lp}}} = \SI{135}{\yocto\newton}$.

\section{Conclusion and outlook}
\label{sec:ConclusionAndOutlook}

In our present setup, fluorescence is detected normal to the plane of the trap chip and, thus, perpendicular to the soft trapping axis along which confinement is almost purely realized by dc fields, labeled as the axial direction of our linear segmented trap (compare \figref{trap_setup}).
In an ideal linear trap, the rf effective potential and thus micromotion depend exclusively on the radial coordinates, perpendicular to the axial direction.
In this case position changes upon rf variation will occur only in radial directions  and thus partially in the direction of our line of sight.
In real traps, there might be a residual axial rf electric field, which is minimized by careful design and typically much smaller than the radial rf field strength. In segmented traps, the intention is usually to exploit the entire axial span of the linear trap by shuttling, etc.,\ and thus axial micromotion might be present but potentially cannot be minimized.
Micromotion minimization always requires a determination of the radial ion displacement or motion, in our case also along our line of sight.
We realize this with focus-scanning imaging as detailed above.
In our present setup this is one of the main limitations, in both precision and speed.
As our detection of excess micromotion is based on a shift of the ion's average position, the method as presented is insensitive to micromotion induced by a phase difference between the rf trap electrodes or an rf pick up of the dc electrodes that is phase-shifted by nearby filter circuits.
These effects can be avoided by design to a great extent, especially in surface traps, and are often of negligible size.

Detection along the axial direction of the trap would allow us to obtain the relevant displacements without the time-consuming physical translation of massive light gathering optics and reduces the number of required images considerably:  right now, we take 25 focal steps and need to include additional times to wait for the completion of a translation action.
Additionally, omitting a translation stage further improves the mechanical stability and thus the precision of the position determination. Taking into account translation and settling times as well as the reduction in the number of images, the speed of acquisition would increase by a factor of approximately 70, meaning that the entire minimization as shown in \figref{compensation2d} would be finished in two dimensions in roughly \SI{9}{\second} and the sensitivity of this method would increase to about \SI{0.3}{\volt\per\meter\sqrt{\hertz}} for a 2D minimization. Speed not only is relevant to keep the time for calibration, etc., low and leave most for the experiment of interest, but also allows to compensate more often and characterize and counteract time-dependent drifts of the power supplies and patch potentials. 
In principle, when the effects of dc field changes of each electrode are well characterized, a single trajectory will suffice, and this might further improve the performance of the method in terms of speed. 

Utilizing the same technique, precise position determination of a harmonically trapped ion may also be used for the simple detection of small forces.
We have demonstrated the detection of forces on the yoctonewton (\si[prefixes-as-symbols=false]{\yocto\newton}) scale in the example of the light pressure force exerted on the ion by the cooling laser. Depending on the experiment, precise knowledge of the light pressure force might be desirable to counter its effects and maintain minimal micromotion also under experimental conditions without incoherent scattering.

In conclusion, we have presented a method, which allows us to compensate micromotion due to dc stray fields in three dimensions using focus-scanning imaging as well as the detection of rather small forces. The method is general and does not require spectroscopy of a narrow transition or a specific orientation of laser beams, which is of particular interest for surface-electrode traps. If imaging is carried out along the symmetry axis of a linear trap and the optimization is carried out in the transverse direction only, no additional components and no focus-scanning imaging are required and the optimization can be carried out within seconds.

\begin{acknowledgments}
We acknowledge funding from the European Community's Seventh
Framework Programme (FP7/2007-2013) under Grant Agreement No. 270843
(iQIT), the European Metrology Research Programm (EMRP) (which is jointly funded by the EMRP participating countries within EURAMET and the European Union),  the Bundesministerium f\"ur Bildung und Forschung (FK 01BQ1012), and from Deutsche Forschungsgemeinschaft.
\end{acknowledgments}


%

\end{document}